# Real-time Measurement of Stress and Damage Evolution During Initial Lithiation of Crystalline Silicon

M. J. Chon,[1] V.A. Sethuraman,[1] A. McCormick,[1] V. Srinivasan,[2] P. R. Guduru[1,*]

[1]School of Engineering, Brown University, 182 Hope Street, Providence, Rhode Island 02912,

[2]Lawrence Berkeley National Laboratory, Berkeley, California 94720

[*]Corresponding author. Email: Pradeep_Guduru@Brown.edu; Telephone: (1) 401 863 3362

**Abstract**

Crystalline to amorphous phase transformation during initial lithiation in (100) silicon-wafers is studied in an electrochemical cell with lithium metal as the counter and reference electrode. It is demonstrated that severe stress jumps across the phase boundary lead to fracture and damage, which is an essential consideration in designing silicon based anodes for lithium ion batteries. During initial lithiation, a moving phase boundary advances into the wafer starting from the surface facing the lithium electrode, transforming crystalline silicon into amorphous $Li_xSi$. The resulting biaxial compressive stress in the amorphous layer is measured *in situ* and it was observed to be *ca*. 0.5 GPa. HRTEM images reveal that the crystalline-amorphous phase boundary is very sharp, with a thickness of ~ 1 nm. Upon delithiation, the stress rapidly reverses, becomes tensile and the amorphous layer begins to deform plastically at around 0.5 GPa. With continued delithiation, the yield stress increases in magnitude, culminating in sudden fracture of the amorphous layer into micro-fragments and the cracks extend into the underlying crystalline silicon.

Silicon is a promising anode material for increasing the energy density of lithium-ion batteries, especially for transportation applications. Since crystalline silicon is the most common and least expensive form of silicon available, many practical attempts to develop silicon-based anodes employ composites that consist of crystalline silicon particles, a softer active phase such as graphite and a binder with a conductive additive [1-4]. Crystalline silicon is also used in more recent anode architectures such as micro/nano wires, nano-porous Si, thin films, nano-particles, *etc* [5-8]. Hence, the structural changes and stresses that arise during initial lithiation and delithiation of crystalline silicon are very important in order to understand and predict damage in silicon-based anodes. Although Si transforms to several distinct Si-Li crystalline phases at high temperatures [9], it has been shown that room temperature electrochemical lithiation of Si results in an amorphous $Li_xSi$ phase, where $x$ is approximately 3.5 [10-13]. It was also shown that, below a potential of about 50 mV *vs.* $Li/Li^+$, amorphous $Li_xSi$ transforms to crystalline $Li_{3.75}Si$ [13]. Upon delithiation, crystalline $Li_{3.75}Si$ is supposed to transform to amorphous $Li_ySi$, where y is between 1 and 2 [13]. However, if the lithiation potential is maintained above ~ 50 mV, $Li_{3.5}Si$ is expected to stay amorphous [13]. Based on these observations, Obrovac and Krause [12] proposed a method to improve the cycling life of silicon-particle based composite anodes by partially transforming crystalline Si to amorphous $Li_xSi$ during initial lithiation and limiting the subsequent lithiation cycles to a potential above the initial lithiation value so that no additional phase transformation takes place. However, the prevailing state of understanding of electrochemically induced phase transformations in crystalline silicon is incomplete without quantitative information on stresses associated with phase transformation and the resulting mechanical damage. Si is known to undergo large volume expansion with upon lithiation (volumetric strain of about 270% in the fully lithiated state). Also, recent experiments of Sethuraman *et al.* [14-16] and Bower *et al.* [17] showed that lithiated silicon can undergo large plastic strains (~ 50%) at a yield stress of the order of 1 GPa. Hence, the large jump in Li concentration across the crystalline-amorphous phase boundary must also involve a large jump in stress and plastic strain. Stresses of the order of 1 GPa have a number of consequences. As argued by Spaepen [18], stress of the order of 1 GPa can modify the thermodynamic driving force for phase transformations. Furthermore, in a Si particle with crystalline core and amorphous shell, compressive stress in the shell induces a tensile stress in the core; as the core shrinks, the tensile stress in it gets amplified, which can result in particle fracture starting from the core. In addition, one of the consequences of plastic deformation during lithiation is that the amorphous shell is subjected to tensile stress during delithiation and it undergoes plastic straining in tension [19]. If the shell is sufficiently thick (*ca.* 100 nm or greater), the tensile stress can result in multiple cracks during delithiation.

The objective of the experiments reported here is to study stress and damage evolution in crystalline silicon during the initial lithiation-delithiation cycle. Double-side polished (100) Si wafers (50.8 mm diameter, 425 – 450 μm thickness) were chosen as working electrodes in an electrochemical cell with Li metal as the counter and reference electrode. A copper grid was fabricated on the wafer side facing the Li foil, which consists of a square array of Cu lines that are 100 nm thick and 5 μm wide; the pitch of the lines is 200 μm in both directions. The Si wafers were then assembled into a specially designed electrochemical cell with a glass window. The Si wafer and the Li counter electrodes separated by a woven Celgard 2500 separator (thickness = 21 μm, Celgard Inc.). 1M $LiPF_6$ in 1:1:1 solution of ethylene carbonate: dimethyl carbonate: diethyl carbonate was used as the electrolyte and wafers were lithiated at a constant current density of 12.5 μA/cm$^2$. A phase boundary moves normal to the surface, transforming the crystalline Si to amorphous lithiated silicon; the simple planar geometry of the phase boundary allows a systematic study of the structural change and the associated stresses. The phase change was seen to occur uniformly under the Cu grid lines as well, and the thickness of the amorphous layer was only about 5% smaller under the Cu lines compared to that away from the Cu lines. It can be concluded that the thin Cu films did not hinder Li diffusion through them at the low current densities used in the experiments. Following lithiation at a constant current for 25 hours, samples were then delithiated at the same flux until the potential reaches the upper limit of 1.2 V *vs*. Li/Li$^+$; subsequently, the potential was held constant until the current falls below 5 nA/cm$^2$.

Stresses in the amorphous layer were measured by monitoring the curvature change in the Si wafer during lithiation and delithiation. The relationship between the biaxial stress in the amorphized region, $\sigma$, and the substrate curvature is given by the Stoney equation [20]. Note that the curvature change is proportional to the product of the thickness of the amorphized layer and the biaxial stress in it. The thickness of the amorphous layer is calculated through mass balance of Li and considering the composition of the amorphous layer to be Li$_{3.5}$Si [13]. Volume expansion ratio of lithiated silicon is taken to be (1+2.7$z$) where $z$ is the state of charge [14]. Wafer curvature was monitored with a multi-beam optical sensor (MOS) wafer curvature system (K-Space Associates); the MOS system uses a parallel array of laser beams that get reflected off the back face of the Si wafer sample and captured on a CCD camera. Wafer curvature is obtained by measuring the relative change in spot spacing [14-16]. In order to image the phase boundary using SEM, samples were removed from the cell, rinsed in dimethyl carbonate, dried and fractured. Sample fragments were then transferred to the SEM chamber. For transmission electron microscopy (TEM) analysis, samples were prepared in a dual beam focused ion beam system (FEI Helios NanoLab 600) and then transferred to the TEM (Jeol 2010) sample stage. It is estimated that the samples were exposed to air for about 5 min during the transfer process. The images are interpreted by assuming that the exposure to air does not alter the sample structure significantly.

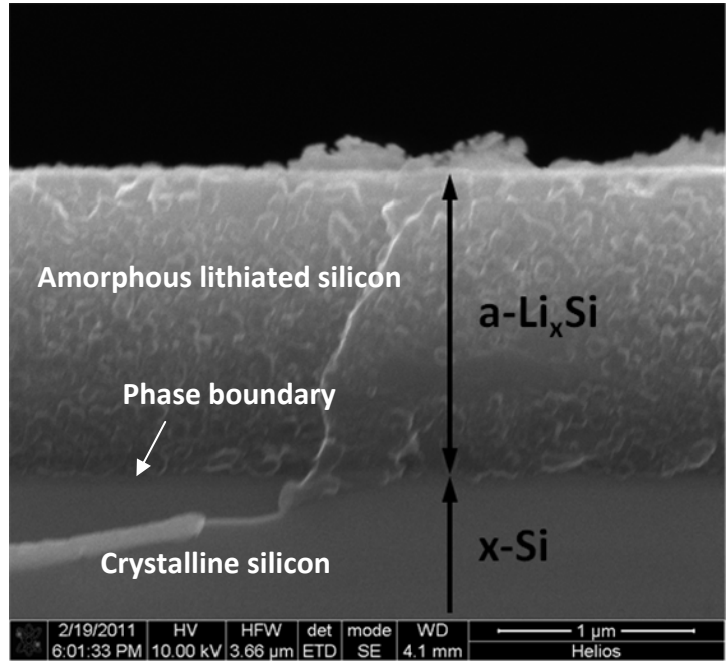

*Figure 1a: Cross-sectional SEM image of the Si wafer following lithiation at a fixed current density of 12.5 µA/cm² for 25 hours. The top layer is amorphous $Li_xSi$, which is separated from the bulk crystalline Si by a sharp phase boundary. Note the change in the morphology of the fracture surface between the two phases.*

A cross-sectional SEM image of the wafer at the end of lithiation is shown in Fig. 1a, revealing the distinct phase boundary between the crystalline Si and the amorphous $Li_xSi$. Also, note the change in the morphology of the fracture surface between the two regions, possibly because crystalline Si is brittle, whereas amorphous $Li_xSi$ can undergo plastic deformation [14]. Fig. 1b shows a TEM image of the phase boundary, which reveals that the phase transition takes place in a very narrow region of ~1 nm width. Figure 2 shows the evolution of cell potential during lithiation; it reaches a plateau, which is the signature of a moving phase boundary. The potential plateau varied between 70 and 115 mV *vs.* Li/Li$^+$ in different samples; the variation is likely due to differences in the cell impedance. These observations are consistent with earlier reports on silicon powder composite anodes [10-13].

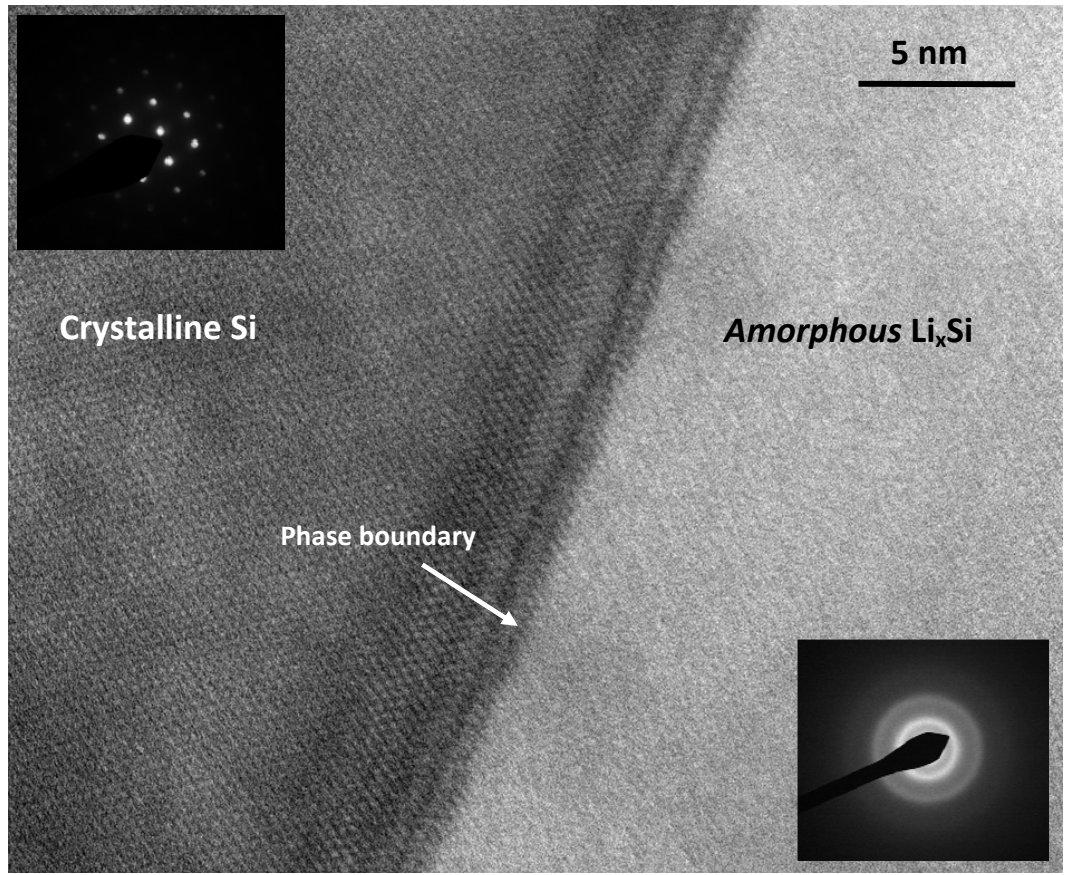

*Figure 1b: HRTEM image of the phase boundary between crystalline Si and amorphous Li$_x$Si, showing a very sharp phase boundary, the width of which is ~ 1 nm. The insets show electron beam diffraction patterns.*

Evolution of the product of stress and thickness of amorphous layer ($\sigma \cdot h$) during lithiation is also shown in Fig. 2, which varies almost linearly with time. Note that the phase boundary speed can be taken to be constant (because the Li flux into the sample is constant), *i.e.*, $h$ increases linearly with time. Consequently, the near linear variation of $\sigma \cdot h$ implies that the biaxial stress in the amorphous layer is approximately constant. Fig. 3 shows stress evolution during the experiment, obtained by dividing the $\sigma \cdot h$ with the amorphous layer thickness. The stress data is noisy initially because both $\sigma \cdot h$ and $h$ start from zero.

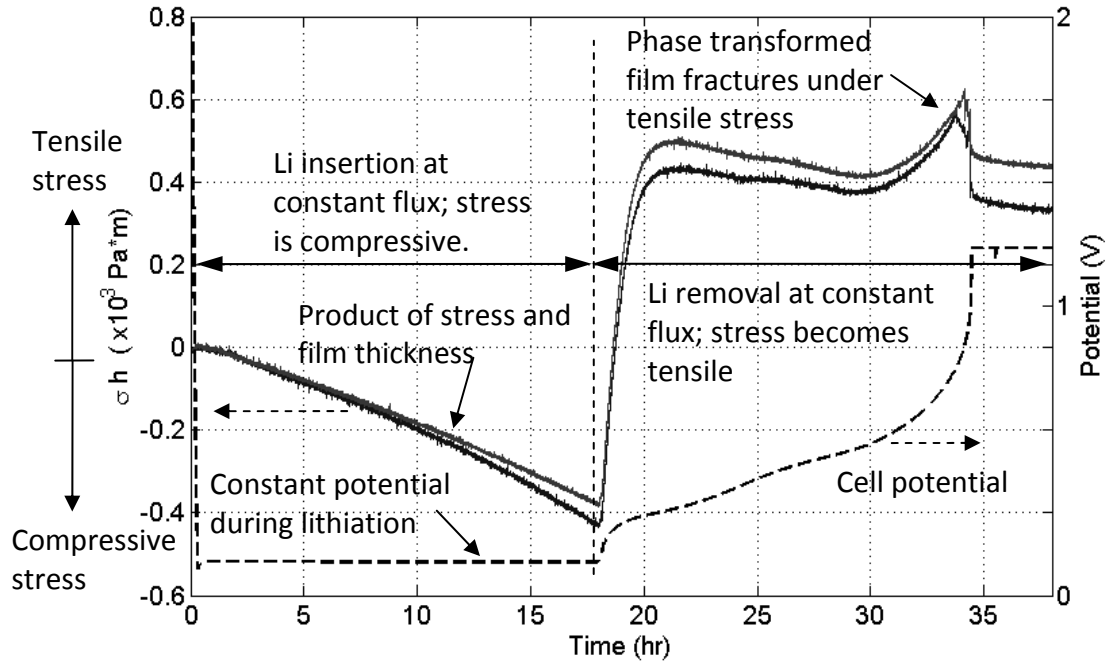

*Figure 2: Evolution of cell potential (dashed) and "stress·thickness, σ·h" (two solid lines corresponding to measurements along two orthogonal directions on the wafer) as a function of time during lithiation and delithiation of a crystalline Si wafer. The potential stays on a plateau during lithiation while a phase boundary moves in the sample, transforming crystalline Si to amorphous lithiated silicon, which is under a state of compressive stress, as indicated by the σ·h plot. During delithiation, the increase in cell potential is similar to that of an amorphous film.*

Averaging over several experiments, the compressive stress in the amorphized layer during lithiation is *ca.* 0.5 GPa (± 0.1 GPa). Since Li concentration undergoes a sharp jump, there is large volume expansion across the phase boundary. However, since the amorphous layer is constrained by the underlying crystalline Si, it must plastically deform biaxially in order to accommodate the volume change. In other words, there is a jump in plastic strain across the phase boundary. The measured stress $\sigma \sim 0.5$ GPa can be viewed as the yield stress of amorphous $Li_{3.5}Si$.

Change in the cell potential and $\sigma \cdot h$ during delithiation are also shown in Fig. 2. Evolution of the cell potential during delithiation is similar to that for amorphous silicon films, as reported in the literature [14,21], which suggests that the phase boundary remains stationary while Li is drained from the amorphous layer.

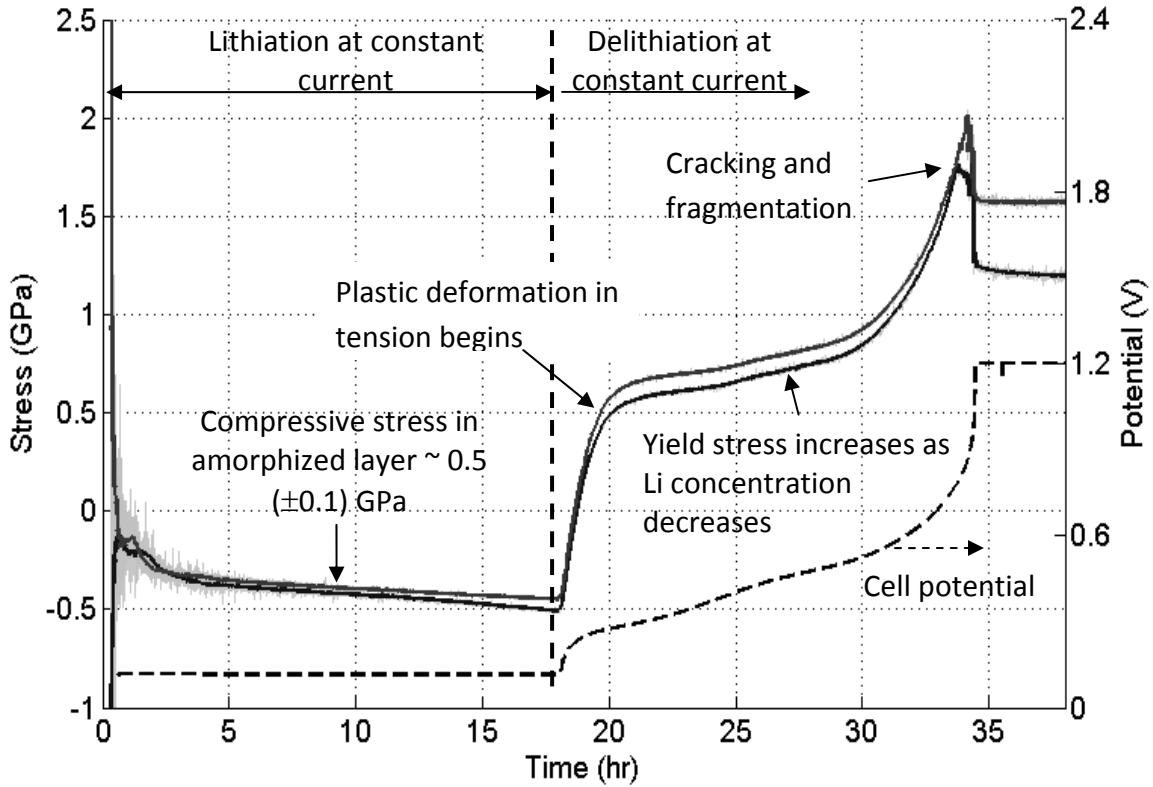

*Figure 3: Evolution of stress in amorphized layer, $\sigma \cdot h$ as a function of time during lithiation and delithiation of a crystalline Si film. (two solid lines corresponding to two orthogonal directions on the wafer). During lithiation, the stress is compressive and nearly constant ~0.5 GPa. During delithiation, the stress rapidly becomes tensile and the film undergoes plastic deformation in tension starting around 0.5 GPa. The yield stress increases beyond 1.5 GPa, resulting in fragmentation of the amorphous layer, as indicated by a sudden drop in stress.*

The stress evolution during delithiation is shown in Fig. 3, which is obtained by considering again that the ratio of the film thickness to that in the fully delithiated state is given by $(1+2.7z)$ [14]. The stress during delithiation rapidly becomes tensile and the amorphous layer begins to deform plastically at a stress of around 0.5 GPa, as indicated by the onset of the stress plateau in Fig. 3. With continued Li removal, the yield stress increases beyond 1.5 GPa. Towards the end of delithiation, the wafer curvature drops rapidly as seen in Fig.2, indicating sudden formation of multiple cracks and fragmentation of the amorphous layer, which is confirmed by SEM imaging of the samples just before and after the sudden drop.

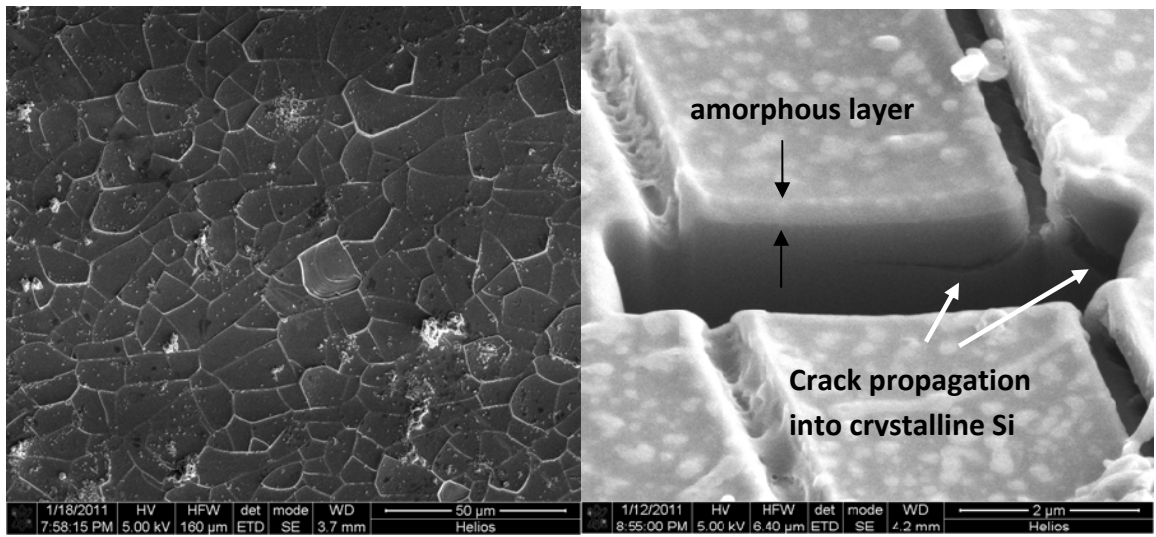

*Figure 4: A network of cracks (left) that forms during delithiation of an amorphized layer. FIB cross section view of the cracks (right) shows that they propagate deeper into the crystalline Si layer, exposing fresh Si surface for SEI formation during subsequent cycles.*

Fig. 4a shows a network of cracks on the surface, which shows the characteristic fragment size to be ~10 μm. In order to investigate the extent of crack propagation into the sample, focused ion beam (FIB) milling was used to create a trench across a crack as shown in Fig. 4b. The cracks were seen to propagate into the underlying crystalline substrate, deeper than just the amorphous layer in which they form. Thus, initial lithiation and delithiation of crystalline silicon results in high compressive and tensile stresses respectively in the amorphized layer, leading to fracture and mechanical damage that stay with the material for the rest of its operational life. These observations are expected to apply to other geometries as well, such as Si particles, nano-wires, etc. if they are partially lithiated. Recently, Rhodes *et al.* [22] monitored acoustic emission (AE) signals from silicon particle composite anodes during lithiation-delithiation cycling and reported high AE activity during the initial lithiation plateau and during delithiation. The former AE activity likely corresponds to tensile fracture at particle cores due to the compressive stress in the outer amorphized layer; the latter corresponds to surface cracking, similar to that shown in Fig. 4. The AE investigation of Rhodes *et al.* [22] shows that the conclusions drawn from idealized wafer geometry extend to particle geometry as well.

Since crystalline silicon is commonly used in silicon-based anodes, it is essential to understand the stresses associated with the crystalline-amorphous phase transformation during initial lithiation-delithiation cycle and the mechanical damage that results. The real-time experimental study on (100) silicon crystals described above reports the following findings: (i) crystalline-amorphous phase transformation during the initial lithiation induces high compressive stress of *ca.* 0.5 GPa in the amorphized layer. Such high compressive stress is unavoidable in all geometries, because of the substrate constraint on the amorphous layer across the moving phase boundary. (ii) TEM imaging of the phase boundary reveals that the phase change occurs over a very narrow region of about 1 nm thickness, which serves as crucial reference data for computational efforts to simulate the amorphization process. (iii) an essential finding is that the stress in the amorphous layer becomes tensile during delithiation and deforms it plastically. The tensile stress exceeds 1.5 GPa as Li concentration decreases and leads to fragmentation of the amorphous layer; the cracks extend beyond the amorphous layer into the underlying crystalline silicon. Consequently, fresh crystalline silicon is exposed to the electrolyte, which can lead to additional SEI (solid electrolyte interphase) formation during the subsequent cycles. At present, silicon anode design is based on a simplistic view of the lithiation-delithiation process in which stress evolution is not considered and the particles are assumed to undergo phase change and volume changes without undergoing damage [12]. The findings reported here contribute important data and observations to improve silicon-based anode design.

The authors gratefully acknowledge the support from NASA EPSCoR (grant #NNX10AN03A), Brown University's NSF-MRSEC program (grant #DMR0079964) and RI Science and Technology Council (grant #RIRA2010-26). V.S. acknowledges support of the Assistant Secretary for Energy Efficiency and Renewable Energy, Office of Vehicle Technologies of the US DoE (contract #DE-AC02-05CH11231).